\documentclass[print,twocolumn]{revtex4}
\usepackage[dvips]{graphicx}
\usepackage{graphicx}
\usepackage{dcolumn}
\usepackage{bm}
\usepackage{epsfig}
\usepackage{latexsym,bm}

\begin{document}

\title{The Reduced properties and applications of the Yangian algebras\\}

\author{Li-Jun TIAN}
\email{tianlijun@shu.edu.cn}
\author{Yan-Ling JIN}
\email{jinyanling@shu.edu.cn}

\affiliation{Department of Physics, Shanghai University, Shanghai,
200444, China\\
Shanghai Key Lab for Astrophysics, Shanghai, 200234, China\\}

\begin{abstract}
The reduced properties and applications of Yangian $Y(sl(2))$ and
$Y(su(3))$ algebras are discussed. By taking a special constraint,
the representation of $Y(su(3))$ can be divided into three 3
$\times$ 3 blocks diagonal based on Gell-mann matrices. The reduced
Yangian $Y(sl(2))$ and $Y(su(3))$ are applied to the bi-qubit system
and the mixed light pseudoscalar meson state respectively, and are
both able to make the final states disentangled after acting on the
initial state by the transition operator, composed of the generators
of Yangian.


PACS number(s): 02.20.-a, 03.65.-w\\
\end{abstract}

\maketitle \baselineskip=12pt

\section{Introduction}

Quantum entanglement, as the nonlocal correlation among different
quantum systems, is of crucial importance in quantum
computation\cite{Nielsen}, quantum teleportation\cite{Bennett1},
dense coding\cite{Bennett2} and quantum key
distribution\cite{Curty}. However, in real systems, the
deterioration of the coherence or even the decoherence and
disentanglement due to the interaction with an environment, which is
recognized as a main obstacle to realize quantum
computing\cite{Beige} and quantum information processing
(QIP)\cite{Viola}, have to be taken into account in the research in
the field of quantum information. Earlier studies had indicated that
entanglement decays exponentially\cite{Yu1, Yu2, Privman} until T.
Yu suggested that entanglement decays completely in finite time and
called for concerted effect to research entanglement sudden
death\cite{Yu}, for example, different systems are
considered\cite{Yonac, Ikram, Cao}, realization in experiment is
given\cite{Almeida} and \cite{Rau} provides theoretical guidance to
practical application of controlling entanglement.

In the last decades, Yangian associated with simple Lie algebras
have been systematically studied in both mathematics and physics and
have many applications through spin operators and quantum fields
\cite{3haldane1}-\cite {3Inozemtsev}. There has been a remarkable
success in studying the long-ranged interaction models by means of
various approaches \cite{3Calogero}-\cite{3Ge2} in which the
Haldane-Shastry model was regarded as the representative of the spin
chain $su(n)$ with long-range interaction
\cite{3Shastry}-\cite{3Bernard}. Recently, Yangian $Y(su(3))$
algebra has been demonstrated to be able to realize the hadronic
decay channels of light pseudoscalar mesons and predict the unknown
particle $X$ in the decay channel
$K_L^0\rightarrow\pi^0\pi^0X$\cite{Tian1}. Moreover, the influence
of transition operators composed of the generators of Yangian
$Y(sl(2))$ and $Y(su(3))$ on the entanglement degrees of two-qubit
system and the mixed light pseudoscalar meson states are discussed
respectively\cite{Tian2, Tian3}.

In this letter, we will follow the similar method in \cite{Tian4},
where the block-diagonal form of Yangian $Y(sl(2))$ algebra is given
by taking a special constraint, to make the $Y(su(3))$ algebra
reduced. And also, an example is presented to compare the effect of
the transition operators of the general Yangian algebras with
reduced ones on the entanglement degrees. Results show that the
generators of the reduced Yangian algebras can make the final states
disentangled while the general ones can't under the same condition.

\section{The Reduced $Y(sl(2))$ Algebra in the Bi-qubit System }

\subsection{The Reduced $Y(sl(2))$ Algebra}

  The Yangian $Y(sl(2))$ is generated by the generators
$\{{I}_{\alpha},{J}_{\alpha}\}$ with the commutation relation
\cite{Ge4}:
\begin{eqnarray}
[{I}_\alpha,{I}_\beta]=i\epsilon_{\alpha\beta\gamma}{I}_\gamma,\;\;\;\;
[{I}_\alpha,{J}_\beta]=i\epsilon_{\alpha\beta\gamma}{J}_\gamma
\;\;\;\;(\alpha,\beta,\gamma=1,2,3),\nonumber
\end{eqnarray}
where the $\{I_{\alpha}\}$ form a simple Lie algebra $ sl(2) $
characterized by $\epsilon_{\alpha\beta\gamma}$ and
\begin{eqnarray}
&&[{J}_{\pm},[{J}_3,{J}_\pm]]=\frac{h^2}{4}{I}_{\pm}
({J}_{\pm}{I}_3-{I}_{\pm}{J}_3),\nonumber
\end{eqnarray}
where $h$ is the deformation parameter and the notations
${I}_\pm={I_1}{\pm}i{I_2}$ and ${J}_\pm={J_1}{\pm}i{J_2}.$

Now let us consider a bi-spin system, in this case the $Y(sl(2))$
generators take the form of\cite{Ge4}
\begin{eqnarray}
{\bf{\emph{\textbf{I}}}}={\bf{\emph{\textbf{S}}}}={\bf{\emph{\textbf{S}}}_1}+{\bf{\textbf{S}}_2},
\end{eqnarray}
\begin{eqnarray}
{\bf{\emph{\textbf{J}}}}=\frac{\mu}{\mu+\nu}{\bf{\emph{\textbf{S}}}}_1}+\frac{\nu}{\mu+\nu}{\bf{\emph{\textbf{S}}_2}+\frac{i\lambda}{\mu+\nu}{\bf{\emph{\textbf{S}}_1}\times{\bf{\emph{\textbf{S}}}}_2},
\end{eqnarray}
where ${\bf{\emph{\textbf{S}}_1}}$, ${\bf{\emph{\textbf{S}}_2}}$ are
the spin-$\frac{1}{2}$ operators and $\mu$, $\nu$ and $\lambda$ are
arbitrary parameters. In this case ${\bf{\emph{\textbf{I}}}}$ is the
total spin operator satisfying $[I_i^a,I_j^b]=i \epsilon_{abc} I_i^c
{\delta}_{i j},(i, j=1, 2)$.

Then direct calculation shows that
\begin{eqnarray}
J^{2}&=&\frac{1}{(\mu+\nu)^2}[\frac{3}{4}({\mu}^2+{\nu}^2-\frac{1}{2}{\lambda}^2)\nonumber\\
&+&(2\mu\nu+\frac{1}{2}\lambda^{2}){S_{1}^{a}}{S_{2}^{a}}],
\end{eqnarray}
and \nonumber\\
\begin{eqnarray}
[J_{a},J_{b}]&=&i\epsilon_{abc}\frac{1}{(\mu+\nu)^2}[-(\mu\nu+\frac{\lambda^2}{4})I_{c}\nonumber\\
&+&(\mu+\nu)^2J_{c}].
\end{eqnarray}
With a special constraint relation \cite{Tian4}
$\mu\nu=-\frac{1}{4}\lambda^{2}$, we can get $J^2=\frac{3}{4}$,
$[J_{a},J_{b}]=i\epsilon_{abc}J_c$. Similarity transformations of
the generators can be made by the use of the matrix $\tau$ who takes
the form of
\begin{eqnarray}
\label{A1} &&\tau=\left(\matrix{1&0&0&0\cr
0&\nu&-{\frac{1}{2}}{\lambda}&0\cr 0&
-{\frac{1}{2}}{\lambda}&\nu&0\cr 0&0&0&1}\right).
\end{eqnarray}
After the similar transformations, the generators become
\begin{eqnarray}
\label{y+1}
&&Y^+=\tau^{-1}J^{+}\tau=\left(\matrix{\frac{\xi{\sigma}^{+}}{2}&0\cr0&\frac{\xi^{-1}{\sigma}^{+}}{2}}\right),\nonumber\\
&&Y^-=\tau^{-1}J^{-}\tau=\left(\matrix{\frac{\xi^{-1}{\sigma}^{-}}{2}&0\cr 0&\frac{\xi{\sigma}^{-}}{2}}\right),\nonumber\\
&&Y^3=\tau^{-1}J^{3}\tau=\frac{1}{2}\left(\matrix{\sigma^3&0\cr \cr 0&\sigma^3}\right),\nonumber\\
\end{eqnarray}
where $\xi$=${\nu}-\frac{1}{2}{\lambda}$ and $\sigma$ are pauli
matrices. \{$Y^{a}$\} reduce to Lie algebra and $4\times4$ matrix is
essentially the 4-dimension representation of $sl(2)$ algebra, so it
is marked as the reduced $Y(sl(2))$ algebra in this case.

\subsection{The Reduced $Y(sl(2))$ Algebra to the Entanglement degree of Bi-qubit System}

For an arbitrary two-qubit pure state
$|\Phi>=\alpha|00>+\beta|01>+\gamma|10>+\delta|11>,$ where
$\alpha,\beta,\gamma,and\;\delta$ are the normalized complex
amplitudes, the concurrence (the entanglement of formation) C is
given by \cite{F, W}
\begin{eqnarray}
\label{6} C=2|\alpha\delta-\beta\gamma|\;\;and\; \;0\leq C\leq1 .
\end{eqnarray}
For a maximally entangled states (MES) $C=1$, we can of course
construct another general state as the initial state
\begin{eqnarray}
|\phi\rangle=\frac{1}{\sqrt2}[\alpha(|00\rangle+|11\rangle)+\beta(|01\rangle+|10\rangle)]
\end{eqnarray}
where $|\alpha|^2+|\beta|^2=1$.

The concurrence C of the initial state is
\begin{eqnarray}
 C=|\alpha^2-\beta^2|.
\end{eqnarray}

If the transition operator is
\begin{eqnarray}
P = a(J^3 + 2 s^3_{1}s^3_{2})
\end{eqnarray}
which is composed of general Yangian $Y(sl(2))$ generators. $a$ is
an arbitrary parameter.

We will get the final state $|\phi^{'}\rangle$
\begin{eqnarray}
|\phi^{'}\rangle=\frac{a}{\sqrt2}[-\frac{\mu-\frac{\lambda}{2}}{\mu+\nu}\beta|01\rangle-\frac{\nu+\frac{\lambda}{2}}{\mu+\nu}\beta|10\rangle+\alpha|11\rangle].
\end{eqnarray}

The normalizing condition is
\begin{eqnarray}
a^2[\alpha^2+\frac{(\mu-\frac{\lambda}{2})^2+(\nu+\frac{\lambda}{2})^2}{(\mu+\nu)^2}\beta^2]=2.
\end{eqnarray}

The concurrence of the final state $|\phi^{'}\rangle$ is
\begin{eqnarray}
\label{c1}
C_\phi^{'}=|\frac{(\mu-\frac{\lambda}{2})(\nu+\frac{\lambda}{2})}{(\mu+\nu)^2}a^2\beta^2|.
\end{eqnarray}

If we add an restrictive condition $\mu=-\nu=\frac{\lambda}{2}$,
namely, it satisfies $\mu\nu=-\frac{\lambda^2}{4}$. So that we can
get the final state $|\phi^{''}\rangle$ acted by reduced Yangian:
\begin{eqnarray}
|\phi^{''}\rangle=\frac{1}{\sqrt2}a\alpha|11\rangle.
\end{eqnarray}

The normalizing condition is
\begin{eqnarray}
a^2\alpha^2=2.
\end{eqnarray}

The concurrence of the final state $|\phi^{''}\rangle$ is
\begin{eqnarray}
\label{c2}
 C_\phi^{''}=0.
\end{eqnarray}

Comparing Eq.(\ref{c1}) with Eq.(\ref{c2}), it's easy to see that
the reduced Yangian $Y(sl(2))$ can make the final state disentangled
while the general one can't.

\section{The Reduced $Y(su(3))$ Algebra in the mixed light pseudoscalar meson states}

\subsection{The Reduced $Y(su(3))$ Algebra}

As is known that the subalgebra of $Y(su(3))$ is Lie algebra $su(3)$
which we have been familiar with the $su(3)$ symmetry for elementary
particles \cite{3GellMann}-\cite{3Ne'eman}. For $su(3)$ generators
is defined by
\begin{eqnarray}
\label{def.f-f}[F^{a},F^{b}]=if_{abc}F^{c}
\end{eqnarray}
where $a$, $b$, $c$ $=$ 1, 2, $\cdots$, 8 and the structure
constants $f_{abc}$ are antisymmetric for any two indices:
\begin{eqnarray}
\label{fabc}
&&f_{123}=1,\;\;\;\;\;\;\;\;\;\;\;\;f_{458}=f_{678}=\frac{\sqrt{3}}{2},\nonumber\\
&&f_{147}=f_{246}=f_{257}=f_{345}=-f_{156}=-f_{367}=\frac{1}{2}.\nonumber\\
\end{eqnarray}
The 3-dimensional representation of $su(3)$ is formed by the
well-known Gell-Mann matrices, i.e.,
\begin{eqnarray}
\label{f}&&{\Lambda}^a=2F^a,\;\;\;\{\Lambda^a,a=1,2,\cdots 8 \}\nonumber\\
&&[\Lambda^{a},\Lambda^{b}]=2if_{abc}\Lambda^{c},
\end{eqnarray}
which eight Hermitian traceless matrices are the extension of Pauli
matrices:
\begin{eqnarray}
\label{Lambda}
\begin{array}{llll}
\vspace{0.3cm}
 \Lambda^1=\left(\matrix{0&1&0\cr 1&0&0\cr 0&
0&0}\right)&\;\;\;\; \Lambda^2=\left(\matrix{0&-i&0\cr i&0&0\cr 0&
0&0}\right)

\cr \vspace{0.3cm}
\Lambda^3=\left(\matrix{1&0&0\cr 0&-1&0\cr 0&
0&0}\right)&\;\;\;\; \Lambda^4=\left(\matrix{0&0&1\cr 0&0&0\cr 1&
0&0}\right)

\cr \vspace{0.3cm}

\Lambda^5=\left(\matrix{0&0&-i\cr 0&0&0\cr i&0&0}\right)&\;\;\;\;
\Lambda^6=\left(\matrix{0&0&0\cr 0&0&1\cr 0&1&0}\right)

\cr\vspace{0.3cm}

\Lambda^7=\left(\matrix{0&0&0\cr 0&0&-i\cr 0&i&0}\right)&\;\;\;\;
\Lambda^8=\frac{1}{\sqrt{3}}\left(\matrix{1&0&0\cr 0&1&0\cr 0&
0&-2}\right)&.
\end{array}
\end{eqnarray}

The antisymmetric structure constants possess the properties:
\begin{eqnarray}
&&f_{ijk}=\frac{1}{4i}Tr([\lambda_{i},\lambda_{j}]\lambda_{k}),\\
&&Tr(\lambda_{i}\lambda_{j})=2\delta_{ij}.
\end{eqnarray}
According to the Jaccobi identity
$[F_{i},[F_{j},F_{k}]]+cycle(i,j,k)=0$, we can get
\begin{equation}
f_{ijm}f_{kmn}+f_{kim}f_{jmn}+f_{jkm}f_{imn}=0.
\end{equation}
Using $Tr([\lambda_{i},\lambda_{j}]\{\lambda_{k},\lambda_{n}\}
+[\lambda_{i},\lambda_{k}]\{\lambda_{j},\lambda_{n}\}
+[\lambda_{i},\lambda_{n}]\{\lambda_{j},\lambda_{k}\})=0$, we have
\begin{equation}
f_{ijm}d_{knm}+f_{ikm}d_{jnm}+f_{inm}d_{jkm}=0.
\end{equation}
There are two Casimir operators in Lie algebra $su(3)$:
\begin{eqnarray}
&&C_{1}=\sum^{8}_{i=1}F^{2}_{i}=-\frac{2i}{3}\sum_{ijk}f_{ijk}F_{i}F_{j}F_{k},\\
&&C_{2}=\sum_{ijk}d_{ijk}F_{i}F_{j}F_{k}=C_{1}(2C_{1}-\frac{11}{6}).
\end{eqnarray}
Introducing the shift operators
\begin{eqnarray}
\label{+-} &&I^{\pm}=F^{1}{\pm}iF^{2},\;\;\;\;\;
U^{\pm}=F^{6}{\pm}iF^{7},\;\;\;\;\;
V^{\pm}=F^{4}{\mp}iF^{5},\nonumber\\
&& I^{3}=F^{3},\;\;\;\;\;\;\;\;\;\;\;\;\;\;
Y=\frac{2}{\sqrt{3}}F^{8}=I^{8},\;\;(Y-hypercharge)\nonumber\\
\end{eqnarray}
the commutation relations (\ref{def.f-f}) can be rewritten in the
form:
\begin{eqnarray}
\label{iuv}
\begin{array}{lll}
\vspace{0.2cm}

 [I^3,I^{\pm}]={\pm}I^{\pm}&\;\;\;\;\;\;
[I^+,I^-]=2I^3,

\cr \vspace{0.2cm}
[I^8,I^{\alpha}]=0(\alpha=\pm,3)&\;\;\;\;\;\;
[I^3,U^{\pm}]={\mp}\frac{1}{2}U^{\pm},

\cr \vspace{0.2cm}
 [I^8,U^{\pm}]={\pm}U^{\pm}&\;\;\;\;\;\;
[U^+,U^{-}]=-I^{3}+\frac{3}{2}I^{8},

\cr \vspace{0.2cm}
[I^3,V^{\pm}]={\mp}\frac{1}{2}V^{\pm}&\;\;\;\;\;\;
[I^8,V^{\pm}]={\mp}V^{\pm},

\cr \vspace{0.2cm}
 [V^+,V^-]=-(I^{3}+\frac{3}{2}I^8)&\;\;\;\;\;\;
[I^{\pm},U^{\pm}]={\pm}V^{\mp},

\cr \vspace{0.2cm}
 [U^\pm,V^{\pm}]={\pm}I^{\mp}&\;\;\;\;\;\;
[V^\pm,I^{\pm}]={\pm}U^{\mp},

\cr \vspace{0.2cm}
 [I^{\pm},U^{\mp}]=0&\;\;\;\;\;\;
 [U^{\pm},V^{\mp}]=0,

\cr \vspace{0.2cm}
 [V^{\pm},I^{\mp}]=0.

\end{array}
\end{eqnarray}
With the notations
\begin{eqnarray}
\label{u3v3} U^3=-\frac{1}{2}I^{3}+\frac{3}{4}I^8,\;\;\;\;
 V^3=-\frac{1}{2}I^{3}-\frac{3}{4}I^8,
\end{eqnarray}
we have
\begin{eqnarray}
\label{uv} &&[U^3,U^{\pm}]={\pm}U^{\pm},\;\;\;\;\;\;\;\;[U^+,U^-]=2U^3,\nonumber\\
&&[V^3,V^{\pm}]={\pm}V^{\pm},\;\;\;\;\;\;\;\;[V^+,V^-]=2V^3.
\end{eqnarray}
Eqs. (\ref{uv}) show that $\{U^{a},a=\pm,3\}$ and
$\{V^{a},a=\pm,3\}$ have the similar commutation relations as the
isospin $\{I^{a},a=\pm,3\}$.
 So, in general, $\vec{U}$ and $\vec{V}$ are called the $U$-spin and $V$-spin.
The charge operator $Q$ is given by the isospin $I^3$ and
hypercharge $Y$ as follows
\begin{eqnarray}
\label{Q}
    Q=I^3+\frac {1}{2}Y.
\end{eqnarray}
And it is easy to check that
\begin{eqnarray}
\label{uq} [U^a,Q]=0\;\;\;\;(a=\pm,3).
\end{eqnarray}
But there is no the same property between $I$-spin and $V$-spin.

For two particles , we define the operators of Y(sl(3)) as follows:
\begin{eqnarray}
\label{def.ij} &&I^a=\sum_iF_i^a,\nonumber\\
  &&J^a=\mu{F_1^a}+\nu{F_2^a}+\frac{i}{2}\lambda{f_{abc}\sum_{i{\neq}j}{\omega}_{ij}F_i^bF_j^c}\;\;(i,j=1,2).\nonumber\\
\end{eqnarray}
Here
\begin{eqnarray}
\label{omega}{\omega}_{ij}=\left\{
\begin{array}{l}
1\;\;\;\;\;\;\;\;i{>}j\\
-1\;\;\;\;\;i{<}j\\
0\;\;\;\;\;\;\;\;i{=}j
\end{array} \right.
\end{eqnarray}
with
$$
{\omega}_{ij}=-{\omega}_{ji}
$$
and $\mu$, $\nu$, $\lambda$ are parameters or casimir operators.
 $\{F_{i}^{a},a=1,2,\cdots,8\}$ form a local $su(3)$ on the $i$
site, and they obey the commutation relation
\begin{eqnarray}
\label{ff} [F_{i}^{a},F_{j}^{b}]=if_{abc}{\delta}_{ij}F_{i}^{c},
\end{eqnarray}
the index $i$ here represents different sites. Substituting
(\ref{def.ij}) into the Yangian commutation relations shown in the
Eqs.(\ref{2 2 i,i}) and (\ref{2 i8i3}), then we can verify the set
$\{{\bf{I},{\bf{J}}}\}$ satisfy $Y(sl(3))$ sufficiently.
Eq.(\ref{def.ij}) plays an important role in explaining the physical
meaning of the representation theory of Chari-Pressley \cite{3chari}
through more calculation \cite{3Bai}

Introducing the notations
\begin{eqnarray}
\label{def.ji+-}
&&{\bar{{I}}^{\pm}}=J^{1}{\pm}iJ^{2},\;\;\;\;{\bar{U}^{\pm}}=J^{6}{\pm}iJ^{7},\;\;\;\;
{\bar{V}^{\pm}}=J^{4}{\pm}iJ^{5},\nonumber\\
&&{\bar{I}^{3}}=J^{3},\;\;\;\;\;\;\;\;\;\;\;\;\;{\bar{I}^{8}}=\frac{2}{\sqrt{3}}J^{8}.
\end{eqnarray}
and from Eqs. (\ref{def.ij}) and (\ref{def.ji+-}) the $Y(sl(3))$ can
be sufficiently realized by
\begin{eqnarray}
\label{jiuv38} I^\pm=&&\sum_iI_i^\pm,\;\;\;\;\;\;U^\pm=\sum_iU_i^\pm,\nonumber\\
V^\pm=&&\sum_iV_i^\pm,\;\;\;\; I^3=\sum_iI_i^3, \;\;\;\;\;\;I^8=\sum_iI_i^8\nonumber\\
{\bar{I}^{\pm}}=&&{\mu}I_{1}^{\pm}+{\nu}I_{2}^{\pm}{\pm}{\lambda}\sum_{i{\neq}j}{\omega}_{ij}
(I_{i}^{\pm}I_{j}^{3}+\frac{1}{2}U_{i}^{\mp}V_{j}^{\mp}),\nonumber\\
{\bar{U}^{\pm}}=&&{\mu}U_{1}^{\pm}+{\nu}U_{2}^{\pm}{\mp}\frac{{\lambda}}{2}\sum_{i{\neq}j}{\omega}_{ij}
[U_{i}^{\pm}(I_{j}^{3}-\frac{3}{2}Y_{j})\nonumber\\
&&+I_{i}^{\mp}V_{j}^{\mp}],\nonumber\\
{\bar{V}^{\pm}}=&&{\mu}V_{1}^{\pm}+{\nu}V_{2}^{\pm}{\mp}\frac{{\lambda}}{2}\sum_{i{\neq}j}{\omega}_{ij}
[V_{i}^{\pm}(I_{j}^{3}+\frac{3}{2}Y_{j})\nonumber\\
&&+U{i}^{\mp}I_{j}^{\mp}],\nonumber\\
{\bar{I}^3}=&&{\mu}I_{1}^{3}+{\nu}I_{2}^{3}-\frac{{\lambda}}{2}\sum_{i{\neq}j}{\omega}_{ij}
[I_{i}^{+}I_{j}^{-}-\frac{1}{2}(U_{i}^{+}U_{j}^{-}\nonumber\\
&&+V_{i}^{+}V_{j}^{-})],\nonumber\\
{\bar{I}^8}=&&{\mu}I_{1}^{8}+{\nu}I_{2}^{8}-\frac{{\lambda}}{2}\sum_{i{\neq}j}{\omega}_{ij}
(U_{i}^{+}U_{j}^{-}-V_{i}^{+}V_{j}^{-}).
\end{eqnarray}
Using Eq.(\ref{iuv}), under the condition of
$\mu\nu=-\frac{{\lambda}^{2}}{4}$ we get
 \begin{equation}
\label{ji+-}
\begin{array}{l}
\begin{array}{ll}
\vspace{0.1cm}
[{\bar{{I}}^{+}},{\bar{{I}}^{-}}]=2(\mu+\nu){\bar{{I}}^{3}}&\;\;\;\;
[{\bar{{I}}^{3}},{\bar{{I}}^{8}}]=0
 \cr\vspace{0.2cm}
[{\bar{{I}}^{3}},{\bar{{I}}^{\pm}}]=\pm(\mu+\nu){\bar{{I}}^{\pm}}&\;\;\;\;
[{\bar{{I}}^{3}},{\bar{{U}}^{\pm}}]=\mp\frac{1}{2}(\mu+\nu){\bar{{U}}^{\pm}}
\cr \vspace{0.2cm}
[{\bar{{I}}^{8}},{\bar{{U}}^{\pm}}]=\pm(\mu+\nu){\bar{{U}}^{\pm}}&\;\;\;\;
[{\bar{{I}}^{3}},{\bar{{V}}^{\pm}}]=\mp\frac{1}{2}(\mu+\nu){\bar{{V}}^{\pm}}
\cr \vspace{0.2cm}
[{\bar{{I}}^{8}},{\bar{{V}}^{\pm}}]=\mp(\mu+\nu){\bar{{V}}^{\pm}}&\;\;\;\;
[{\bar{{U}}^{+}},{\bar{{U}}^{-}}]=(\mu+\nu)(-{\bar{{I}}^{3}}+\frac{3}{2}{\bar{{I}}^{8}})
\cr \vspace{0.2cm}
[{\bar{{I}}^{\pm}},{\bar{{U}}^{\pm}}]=\pm(\mu+\nu){\bar{{V}}^{\mp}}&\;\;\;\;
[{\bar{{V}}^{+}},{\bar{{V}}^{-}}]=(\mu+\nu)(-{\bar{{I}}^{3}}-\frac{3}{2}\bar{{{I}}^{8}})
\cr \vspace{0.2cm}
[{\bar{{V}}^{\pm}},{\bar{{I}}^{\pm}}]=\pm(\mu+\nu){\bar{{U}}^{\mp}}&\;\;\;\;
[{\bar{{U}}^{\pm}},{\bar{{V}}^{\pm}}]=\pm(\mu+\nu){\bar{{I}}^{\mp}}
\cr \vspace{0.2cm}
[{\bar{{I}}^{\pm}},{\bar{{U}}^{\pm}}]={\pm}{\bar{{V}}^{\mp}}&\;\;\;\;
[{\bar{{V}}^{\mp}},{\bar{{I}}^{\pm}}]=[{\bar{{U}}^{\pm}},{\bar{{V}}^{\mp}}]
=[{\bar{{I}}^{\mp}},{\bar{{U}}^{\pm}}]\nonumber\\&\;\;\;\;=0.

\end{array}
\end{array}
\end{equation}
If we introduce the notations
\begin{eqnarray}
\label{notation}
\bar{U}^3=-\frac{1}{2}\bar{I}^{3}+\frac{3}{4}\bar{I}^8,\;\;\;\;\;\;\;\;\;\;\;\;
 \bar{V}^3=-\frac{1}{2}\bar{I}^{3}-\frac{3}{4}\bar{I}^8,
\end{eqnarray}
then
$$
[{\bar{{U}}^{+}},{\bar{{U}}^{-}}]=2(\mu+\nu){\bar{{U}}^{3}},\;\;\;\;
[{\bar{{V}}^{+}},{\bar{{V}}^{-}}]=2(\mu+\nu){\bar{{V}}^{3}}.
$$
With the help of Eq.(\ref{jiuv38}) and the condition of
(\ref{munulambda}), direct calculation shows
\begin{eqnarray}
\label{j2}
({\bf{J}})^2=&&(J^{1})^{2}+(J^{2})^{2}+(J^{3})^{2}+(J^{4})^{2}+(J^{5})^{2}\nonumber\\
&&+(J^{6})^{2}+(J^{7})^{2}+(J^{8})^{2}\nonumber\\
=&&(J^{I^+})^{2}+(J^{I^-})^{2}+(J^{I^3})^{2}+(J^{U^+})^{2}+(J^{U^-})^{2}\nonumber\\
&&+(J^{V^+})^{2}+(J^{V^{-}})^{2}+(\frac{\sqrt{3}}{2}J^{I^8})^{2}\nonumber\\
=&&\frac{1}{3}(\mu+\nu)^2
\end{eqnarray}
It means that if we set
\begin{eqnarray}
\label{y} Y^{a}=\frac{1}{\mu+\nu}J^{a},\;\;\;\;(a=1,2,\cdots,8)
\end{eqnarray}
in terms of the notations
\begin{eqnarray}
\label{notation1}
&&\tilde{I}^{\pm}={Y}^{1}{\pm}iY^{2},\;\;\;\;{\tilde{{U}}^{\pm}}=Y^{6}{\pm}iY^{7},\;\;\;\;
{\tilde{{V}}^{\pm}}=Y^{4}{\pm}iY^{5},\nonumber\\
&&{\tilde{I}^{3}}=Y^{3},\;\;\;\;\;\;\;\;\;\;\;\;\;{\tilde{I}^{8}}=\frac{2}{\sqrt{3}}Y^{8},
\end{eqnarray}
we have
\begin{eqnarray}
\label{y2} ({\bf{Y}})^2=\frac{1}{3}.
\end{eqnarray}
In the following we get the commutation relations
\begin{equation}
\label{breveI}
\begin{array}{l}
\begin{array}{lll}
\vspace{0.2cm}
[{\tilde{I}^{+}},{\tilde{I}^{-}}]=2{\tilde{I}^{3}}&\;\;
[{\tilde{{I}}^{3}},{\tilde{{I}}^{8}}]=[{\tilde{{I}}^{\pm}},{\tilde{{I}}^{8}}]=0

\cr \vspace{0.2cm}
[{\tilde{{I}}^{3}},{\tilde{I}^{\pm}}]={\pm}{\tilde{{I}}^{\pm}}&\;\;
[{\tilde{{I}}^{3}},{\tilde{U}^{\pm}}]={\mp}\frac{1}{2}{\tilde{U}^{\pm}}

\cr \vspace{0.2cm}
[\tilde{{I}}^{8},{\tilde{U}^{\pm}}]={\pm}{\tilde{U}^{\pm}}&\;\;
[{\tilde{I}^{3}},{\tilde{V}^{\pm}}]={\mp}\frac{1}{2}{\tilde{V}^{\pm}}

\cr \vspace{0.2cm}
[\tilde{I}^{8},\tilde{V}^{\pm}]={\mp}{\tilde{V}^{\pm}}&\;\;
[{\tilde{U}^{+}},{\tilde{U}^{-}}]=2\tilde{U}^3

\cr \vspace{0.2cm}
[{\tilde{I}^{\pm}},{\tilde{U}^{\pm}}]={\pm}{\tilde{V}^{\mp}}&\;\;
[{\tilde{V}^{+}},{\tilde{V}^{-}}]=2\tilde{V}^3

\cr \vspace{0.2cm}
[{\tilde{V}^{\pm}},{\tilde{I}^{\pm}}]={\pm}{\tilde{U}^{\mp}}&\;\;
[{\tilde{U}^{\pm}},{\tilde{V}^{\pm}}]={\pm}{\tilde{I}^{\mp}}

 \cr \vspace{0.2cm}
[{\tilde{I}^{\pm}},{\tilde{U}^{\pm}}]={\pm}{\tilde{V}^{\mp}}&\;\;
[{\tilde{V}^{\mp}},{\tilde{I}^{\pm}}]=[{\tilde{U}^{\pm}},{\tilde{V}^{\mp}}]=0

\cr \vspace{0.2cm} [{\tilde{I}^{\mp}},{\tilde{U}^{\pm}}]=0,
\end{array}
\end{array}
\end{equation}
where $\tilde{U}^3$ =
$-\frac{1}{2}\tilde{I}^{3}+\frac{3}{4}\tilde{I}^8$ and $
\tilde{V}^3$ = $-\frac{1}{2}\tilde{I}^{3}-\frac{3}{4}\tilde{I}^8$.

It is similar with the commutation relations of the $I$-spin,
$U$-spin, and $V$-spin, namely the Equation (\ref{iuv}). By the
discussion we have the result: a general realization of $Y(su(3))$
is given by equations
\begin{eqnarray}
\label{ya}
I^a=&&{\sum_i}F_i^a,\nonumber\\
Y^a=&&\frac{\mu}{\mu+\nu}I_{1}^{a}+\frac{\nu}{\mu+\nu}I_{2}^{a}+\frac{i\lambda}{2(\mu+\nu)}
f_{abc}\sum_{i\neq j}{\omega}_{ij}I_i^{b}I_j^{c}\nonumber\\
&&(i,j=1,2).
\end{eqnarray}
Then if we take the condition $\mu\nu$=-$\frac{{\lambda}^2}{4}$ and
the fundamental representation of local $su(3)$ is held, they have
the same commutation relations between the two generators of
Yangian, namely $Y(su(3))$ algebra has reduced to two sets of
$su(3)$ algebras.

In fact, the fundamental representation of local $su(3)$ is given by
\begin{equation}
\label{iuv38}
\begin{array}{l}
\begin{array}{lll}
\vspace{0.3cm} I^{+}=\left( \matrix{0&1&0 \cr 0&0&0 \cr
0&0&0}\right)&\;\;\; U^{+}=\left( \matrix{0&0&0 \cr 0&0&1 \cr
0&0&0}\right) \cr V^{+}=\left( \matrix{0&0&0 \cr 0&0&0 \cr
1&0&0}\right)&\;\;\; I^{3}=\frac{1}{2}\left( \matrix{1&0&0 \cr
0&-1&0 \cr \vspace{0.3cm}
 0&0&0}\right) \cr
 Y=\frac{1}{3}\left( \matrix{1&0&0 \cr 0&1&0
\cr 0&0&-2}\right),
\end{array}
\end{array}
\end{equation}
and $I^{-}$=$(I^{+})^{+}$, $U^{-}$=$(U^{+})^{+}$,
$V^{-}=(V^{+})^{+}$.

By using Eqs.(\ref{jiuv38}), (\ref{y}) and (\ref{iuv38}), we get
\begin{eqnarray}
{\tilde{I}^+}=&&\frac{1}{\mu+\nu}[{\mu}I_{1}^{+}+{\nu}I_{2}^{+}+{\lambda}(I_1^{+}I_2^{3}
+\frac{1}{2}U_1^{-}V_2^{-})\nonumber\\
&&-{\lambda}(I_1^{3}I_2^{+} +\frac{1}{2}V_1^{-}U_2^{-})]\nonumber\\
{\tilde{{I}}^-}=&&\frac{1}{\mu+\nu}[{\mu}I_{1}^{-}+{\nu}I_{2}^{-}-{\lambda}(I_1^{-}I_2^{3}
+\frac{1}{2}U_1^{+}V_2^{+})\nonumber\\
&&+{\lambda}(I_1^{3}I_2^{-}
+\frac{1}{2}V_1^{+}U_2^{+})]\nonumber\\
{\tilde{{U}}^+}=&&\frac{1}{\mu+\nu}\{{\mu}U_{1}^{+}+{\nu}U_{2}^{+}-\frac{{\lambda}}{2}[U_1^{+}
(I_2^{3}-\frac{3}{2}Y_{2})\nonumber\\
&&+I_1^-{V_{2}^-}-U_2^{+}(I_1^{3}-\frac{3}{2}Y_{1})I_2^{-}V_1^{-}]\}\nonumber\\
{\tilde{{U}}^-}=&&\frac{1}{\mu+\nu}\{{\mu}U_{1}^{-}+{\nu}U_{2}^{-}+\frac{{\lambda}}{2}[U_1^{-}
(I_2^{3}-\frac{3}{2}Y_{2})\nonumber\\
&&+I_1^+V_{2}^{+}-(I_1^{3}-\frac{3}{2}Y_{1})U_2^{-}+V_1^{+}I_2^{+}]\}\nonumber\\
{\tilde{{V}}^+}=&&\frac{1}{\mu+\nu}\{{\mu}V_{1}^{+}+{\nu}V_{2}^{+}-\frac{{\lambda}}{2}[V_1^{+}
(I_2^{3}+\frac{3}{2}Y_{2})\nonumber\\
&&+U_1^-{V_{2}^-}-(I_1^{3}+\frac{3}{2}Y_{1})V_2^{+}-I_1^{-}U_2^{-}]\}\nonumber\\
{\tilde{V}^-}=&&\frac{1}{\mu+\nu}\{{\mu}V_{1}^{-}+{\nu}V_{2}^{-}+\frac{{\lambda}}{2}[V_1^{-}
(I_2^{3}+\frac{3}{2}Y_{2})\nonumber\\
&&+U_{1}^+{I_2^+}-(I_1^{3}+\frac{3}{2}Y_{1})V_2^{-}-I_1^{+}U_2^{+}]\}\nonumber\\
{\tilde{{I}}^3}=&&\frac{1}{\mu+\nu}\{{\mu}I_{1}^{3}+{\nu}I_{2}^{3}-\frac{{\lambda}}{2}[I_1^{+}I_2^{-}
-\frac{1}{2}(U_{1}^{+}U_{2}^{-}\nonumber\\
&&+V_{1}^{+}V_{2}^{-})-I_1^{-}I_2^{+}+\frac{1}{2}(U_{1}^{-}U_2^{+}+V_1^{-}V_2^{+})]\}\nonumber\\
{\tilde{{I}}^8}=&&\frac{1}{\mu+\nu}[{\mu}I_{1}^{8}+{\nu}I_{2}^{8}
-\frac{{\lambda}}{2}(U_{1}^{+}U_{2}^{-}\nonumber\\
&&-V_{1}^{+}V_{2}^{-})+\frac{{\lambda}}{2}(U_{1}^{-}U_2^{+}-V_1^{-}V_2^{+})]
\end{eqnarray}

Setting $x$ the eigenvalue of $\tilde{I}^3$, then
$|xE-\tilde{I}^3|=0$ with $E$ a unite matrix and its solutions are
\begin{eqnarray}
\label{x2}
&&x_1=0,\;\;\;\;x_{2,3}=\pm\frac{1}{2},\nonumber\\
&&x_{4,5}=\pm\frac{1}{2(\mu+\nu)}\sqrt{\mu^2-2\mu\nu+\nu^2-\lambda^2},\nonumber\\
&&x_{6,7}=\frac{1}{4}\pm\frac{1}{4(\mu+\nu)}\sqrt{\mu^2-2\mu\nu+\nu^2-\lambda^2}\nonumber\\
&&x_{8,9}=-\frac{1}{4}\pm\frac{1}{4(\mu+\nu)}\sqrt{\mu^2-2\mu\nu+\nu^2-\lambda^2}.
\end{eqnarray}
Taking $\mu\nu=-\frac{\lambda^2}{4}$, then we can get
\begin{eqnarray}
\label{x}
x_{1,2,3}=0,\;\;\;\;x_{4,5,6}=\frac{1}{2},\;\;\;\;x_{7,8,9}=-\frac{1}{2}.
\end{eqnarray}
If we take the similar matrix as
\begin{eqnarray}
\label{su3a} A=\left(\matrix{1&0&0&0&0&0&0&0&0 \cr
0&\nu&0&-\frac{\lambda}{2}&0&0&0&0&0 \cr
0&0&\nu&0&0&0&-\frac{\lambda}{2}&0&0 \cr
0&-\frac{\lambda}{2}&0&\nu&0&0&0&0&0 \cr 0&0&0&0&1&0&0&0&0 \cr
0&0&0&0&0&\nu&0&-\frac{\lambda}{2}&0 \cr
0&0&-\frac{\lambda}{2}&0&0&0&\nu&0&0 \cr
0&0&0&0&0&-\frac{\lambda}{2}&0&\nu&0 \cr 0&0&0&0&0&0&0&0&1}\right),
\end{eqnarray}
and its inverse matrix $A^{-1}$ can be obtained easily, thus we get
\begin{equation}
\label{iuv38'}
\begin{array}{ll}
\vspace{0.4cm} ({\tilde{I}^{3}})^{'}=A^{-1}{\tilde{I}^{3}}A=\left(
\matrix{{I}^{3}&0&0 \cr 0&I^{3}&0 \cr 0&0&I^{3}}\right)&

\cr \vspace{0.4cm}

({\tilde{I}^{8}})^{'}=A^{-1}{\tilde{I}^{8}}A=\left(
\matrix{I^{8}&0&0 \cr 0&I^{3}&0 \cr 0&0&I^{8}}\right)

\cr \vspace{0.4cm}

({\tilde{I}^{+}})^{'}=A^{-1}{\tilde{I}^{+}}A=\left(
\matrix{{\alpha}I^{+}&0&0 \cr 0&{\alpha}^{-1}I^{+}&0 \cr
0&0&I^{+}}\right)&

\cr \vspace{0.4cm}

 ({\tilde{I}^{-}})^{'}=A^{-1}{\tilde{I}^{-}}A=\left(
\matrix{{\alpha}^{-}I^{-}&0&0 \cr 0&{\alpha}I^{-}&0 \cr
0&0&I^{-}}\right)

\cr  \vspace{0.4cm}

 ({\tilde{U}^{+}})^{'}=A^{-1}{\tilde{U}^{+}}A=\left(
\matrix{U^{+}&0&0 \cr 0&{\alpha}U^{+}&0 \cr
0&0&{\alpha}^{-}U^{+}}\right)&

\cr \vspace{0.4cm}

({\tilde{U}^{-}})^{'}=A^{-1}{\tilde{U}^{-}}A=\left( \matrix{
U^{-}&0&0 \cr 0&{\alpha}^{-1}U^{-}&0 \cr 0&0&{\alpha}U^{-}}\right)

\cr  \vspace{0.4cm}

({\tilde{V}^{+}})^{'}=A^{-1}{\tilde{V}^{+}}A=\left(
\matrix{{\alpha}^{-}V^{+}&0&0 \cr 0&V^{+}&0 \cr
0&0&{\alpha}V^{+}}\right)&

\cr \vspace{0.4cm}

({\tilde{V}^{-}})^{'}=A^{-1}{\tilde{V}^{-}}A=\left( \matrix{
{\alpha}V^{-}&0&0 \cr 0&V^{-}&0 \cr 0&0&{\alpha}^{-1}V^{-}}\right)

\end{array}
\end{equation}

where $\alpha=\nu-\frac{\lambda}{2}$. So in virtue of a similar
transformation, we reduce the eight 9 by 9 matrix to the three 3 by
3 block diagonal.

Taking the correspondence,
\begin{eqnarray}
\label{y'} &&I^{+}{ \rightarrow}{\alpha}I^{+},\;\;\;I^{-}{
\rightarrow}{\alpha}^{-1}I^{-},\;\;\;
I^{3}{ \rightarrow}I^{3},\;\;\;I^{8}{ \rightarrow}I^{8},\nonumber\\
&&U^{+}{ \rightarrow}{\alpha}U^{+},\;\;U^{-}{
\rightarrow}{\alpha}^{-1}U^{-},\;\; V^{+}{
\rightarrow}{\alpha}V^{+},\;\;V^{-}{
\rightarrow}{\alpha}^{-1}V^{-},\nonumber
\end{eqnarray}
Eq.(\ref{breveI}) will get the same result, that is, the Yangian
algebra we discussed hides a $u(1)$ algebra.

\subsection{The Reduced $Y(su(3))$ Algebra to the entanglement degree of the mixed light pseudoscalar meson states}

The initial state is
\begin{eqnarray}
|\varphi\rangle=\alpha_1|\kappa^+\rangle+\alpha_2|\kappa^-\rangle,
\end{eqnarray}
where $\alpha_1$ and $\alpha_2$ are the normalized real amplitudes
and they satisfy $\alpha_1^2+\alpha_2^2=1$.
$|\kappa^+\rangle=|u\bar{s}\rangle$,
$|\kappa^-\rangle=|s\bar{u}\rangle$.

The degree of entanglement of the initial state $|\varphi\rangle$ is
\begin{eqnarray}
C_\varphi=-\alpha_1^2Log_3\alpha_1^2-\alpha_2^2Log_3\alpha_2^2.
\end{eqnarray}

If the transition operator is
\begin{eqnarray}
P=\eta_1\bar{V}^++\eta_2\bar{V}^-
\end{eqnarray}
which is composed of general Yangian generators.

We will get the final state $|\varphi^{'}\rangle$
\begin{eqnarray}
|\varphi^{'}\rangle=&&\frac{1}{\sqrt3}[(\mu+\nu)\eta_2\alpha_1+(\mu+\lambda)\eta_2\alpha_2\nonumber\\
&&+(\mu+\nu)\eta_1(\alpha_1+\alpha_2)]|\eta^{0'}\rangle\nonumber\\
&&+\frac{1}{\sqrt2}[(\mu+\nu)\eta_2\alpha_1+(\mu+\lambda)\eta_2\alpha_2]|\pi^0\rangle\nonumber\\
&&+\frac{1}{\sqrt6}[2(\mu+\nu)\eta_1(\alpha_1+\alpha_2)-(\mu+\nu)\eta_2\alpha_1\nonumber\\
&&-(\mu+\lambda)\eta_2\alpha_2]|\eta^0\rangle.
\end{eqnarray}

The normalizing condition is
\begin{eqnarray}
[(\mu+\nu)\eta_2\alpha_1+(\mu+\lambda)\eta_2\alpha_2]^2+(\mu+\nu)^2\eta_1^2(\alpha_1+\alpha_2)^2=1.
\end{eqnarray}

The degree of entanglement of the final state $|\varphi^{'}\rangle$
is
\begin{eqnarray}
\label{c3}
C_\varphi^{'}=&&-[(\mu+\nu)\eta_2\alpha_1+(\mu+\lambda)\eta_2\alpha_2]^2Log_3[(\mu+\nu)\eta_2\alpha_1\nonumber\\
&&+(\mu+\lambda)\eta_2\alpha_2]^2-(\mu+\nu)^2\eta_1^2(\alpha_1+\alpha_2)^2\nonumber\\
&&Log_3(\mu+\nu)^2\eta_1^2(\alpha_1+\alpha_2)^2.
\end{eqnarray}

If we add an restrictive condition $\mu=-\nu=\frac{\lambda}{2}$,
namely, it satisfies $\mu\nu=-\frac{\lambda^2}{4}$. So that we can
get the final state $|\varphi^{''}\rangle$ acted by reduced Yangian:
\begin{eqnarray}
|\varphi^{''}\rangle=\frac32\lambda\eta_2\alpha_2(\frac{1}{\sqrt3}|\eta^{0'}\rangle+\frac{1}{\sqrt2}|\pi^0\rangle-\frac{1}{\sqrt6}|\eta^0\rangle)
\end{eqnarray}

The normalizing condition is
\begin{eqnarray}
\frac94\lambda^2\eta_2^2\alpha_2^2=1.
\end{eqnarray}

The degree of entanglement of the final state $|\varphi^{''}\rangle$
is
\begin{eqnarray}
\label{c4}
 C_\varphi^{''}=0.
\end{eqnarray}

We can obtain the same conclusion that the reduced Yangian
$Y(su(3))$ can make the final state disentangled while the general
one can't from Eq.(\ref{c3}) and Eq.(\ref{c4}).

\section{Conclusion}

In this paper, we have discussed the reduced properties and
applications of Yangian $Y(sl(2))$ and $Y(su(3))$. As same as
$Y(sl(2))$, $Y(su(3))$ algebra has been reduced into two sets of
$su(3)$ algebras, moreover the structure is the same as $su(2)$
case, i.e. those formed by the generators ${\bf{Y}}$ of Yangian are
all constructed by the consequent generators of $su(3)$. Moreover,
we have compared the influence of the transition operators of
general Yangian algebras with reduced ones on the entanglement
degrees and found that the reduced ones can make the initial states
disentangled while the general ones can not both for $Y(sl(2))$ and
$Y(su(3))$.

Now a question put forward: can we use this method to $su(n)$? To
our knowledge, this problem has not been discussed in this thesis.
But we can surmise that for $Y(su(n))$ the matrices of the
generators ${\bf{Y}}$ can be written as $n$ pieces of n $\times$ n
matrices, furthermore each pieces is formed by the consequently
generators of $su(n)$. Consequently, it is of interesting area how
to generalize the idea in $Y(su(n))$, which makes the system contact
with physical application.

\section{Acknowledgements} This work is in part supported by the NSF of
China under Grant No. 10775092 and No.10875026, Shanghai Leading
Academic Discipline Project (Project number S30105) and Shanghai
Research Foundation No.07d222020.

\end{document}